\documentclass{article}

\usepackage{arxiv}

\usepackage[utf8]{inputenc} 
\usepackage[T1]{fontenc}    
\usepackage{hyperref}       
\usepackage{url}            
\usepackage{booktabs}       
\usepackage{amsfonts}       
\usepackage{nicefrac}       
\usepackage{microtype}      
\usepackage{lipsum}
\usepackage{graphicx}
\graphicspath{ {./images/} }

\usepackage{tabularx} 
\usepackage{makecell} 
\usepackage{booktabs} 
\usepackage{multirow} 
\usepackage{hyperref}
\usepackage{cleveref}
\usepackage{array} 
\usepackage{graphicx}

\usepackage{listings}
\usepackage{fancyvrb}
\usepackage{rotating}
\usepackage{longtable}
\usepackage{pdflscape} 
\usepackage{algorithm}
\usepackage{algpseudocode}
\usepackage[table]{xcolor}
\usepackage{comment}
\usepackage{soul}

\title{CICAPT-IIoT: A Provenance-Based APT Attack Dataset for  IIoT Environment}

\author{
  Erfan Ghiasvand \\
  Faculty of Computer Science\\
  University of New Brunswick\\
  Fredericton, NB, Canada\\
  \texttt{eghiasva@unb.ca} \\
   \And
  Suprio Ray \\
  Faculty of Computer Science\\
  University of New Brunswick\\
  Fredericton, NB, Canada\\
  \texttt{sray@unb.ca} \\
   \And
  Shahrear Iqbal \\
  National Research Council\\
  Fredericton, NB, Canada\\
  \texttt{shahrear.iqbal@nrc-cnrc.gc.ca} \\
   \And
  Sajjad Dadkhah \\
  Faculty of Computer Science\\
  University of New Brunswick\\
  Fredericton, NB, Canada\\
  \texttt{sdadkhah@unb.ca} \\
  \And
  Ali A. Ghorbani \\
  Faculty of Computer Science\\
  University of New Brunswick\\
  Fredericton, NB, Canada\\
  \texttt{ghorbani@unb.ca} \\
}

\begin{document}
\maketitle
\begin{abstract}
The Industrial Internet of Things (IIoT) is a transformative paradigm that integrates smart sensors, advanced analytics, and robust connectivity within industrial processes, enabling real-time data-driven decision-making and enhancing operational efficiency across diverse sectors, including manufacturing, energy, and logistics. IIoT is susceptible to various attack vectors, with Advanced Persistent Threats (APTs) posing a particularly grave concern due to their stealthy, prolonged, and targeted nature. The effectiveness of machine learning-based intrusion detection systems in APT detection has been documented in the literature. However, existing cybersecurity datasets often lack crucial attributes for APT detection in IIoT environments. 
Incorporating insights from prior research on APT detection using provenance data and intrusion detection within IoT systems, we present the CICAPT-IIoT dataset. The main goal of this paper is to propose a novel APT dataset in the IIoT setting that includes essential information for the APT detection task. In order to achieve this, a testbed for IIoT is developed, and over 20 attack techniques frequently used in APT campaigns are included. The performed attacks create some of the invariant phases of the APT cycle, including Data Collection and Exfiltration, Discovery and Lateral Movement,  Defense Evasion, and Persistence. By integrating network logs and provenance logs with detailed attack information, the CICAPT-IIoT dataset presents foundation for developing holistic cybersecurity measures. Additionally, a comprehensive dataset analysis is provided, presenting cybersecurity experts with a strong basis on which to build innovative and efficient security solutions.
\end{abstract}


\section{Introduction}
Advanced Persistent Threats (APTs) represent a sophisticated category of cyberattacks, where an unauthorized user gains access to a network and remains undetected for a long period of time. Some attackers aim to harm organizations for financial motives or to gain notoriety by damaging a company's reputation, and they do not conceal their actions. However, in recent years, another type of attacker group has risen in prominence, which is characterized by a deliberate and methodical approach. They employ a ``low and slow'' strategy with the goal of either stealing sensitive data from their targets or disrupting their operations \cite{alshamrani2019survey}. APTs represent a significant threat to critical infrastructure systems and have been responsible for numerous severe incidents. APT attacks are distinguished from typical cyberattacks by some key characteristics, such as complexity, persistence, being targeted, and elusiveness. APT attacks typically consist of several distinct phases, each with specific objectives and strategies. While the exact phases can vary depending on the attack group and campaign, the following \cite{milajerdi2019holmes} are common phases in an APT attack: (1) Initial Compromise, (2) Establishing a Foothold, (3) Privilege Escalation, (4) Reconnaissance, (5) Lateral Movement, (6) Maintaining Persistence, and (7) Data Collection and Exfiltration. \\
Indeed, Industrial Internet of Things (IIoT) networks represent a particularly vulnerable target for APT attacks. Originally centered around general applications, IoT has extended its influence to diverse sectors, including industry, where there is an increasing drive to interconnect previously isolated components, facilitating both intra-component communication and connections to the broader Internet \cite{malik2021industrial}. Industrial IoT enables the seamless integration of several devices with sensing, identification, processing, communication, and networking capabilities \cite{da2014internet}. Researchers use many system architectures for IIoT systems, such as Brown-IIoTbed~\cite{al2020developing}, to develop an IIoT environment. \\
The security and safety of IIoT systems have been the subject of substantial research due to the essential importance and sensitivity of Industrial IoT applications. As demonstrated by historical occurrences like Stuxnet \cite{langner2011stuxnet}, the Ukrainian power plant attacks \cite{whitehead2017ukraine}, and the TRITON incident \cite{di2018triton}, attacks on the IIoT can have significant consequences that go beyond the scope of a company's operations and may compromise the safety of citizens and even the entire nation. 
Research findings on the security of Industrial IoT reveal the disturbing fact that IIoT devices are susceptible to weaknesses, as described in \cite{wurm2016security} and \cite{sisinni2018industrial}. This paints an alarming picture of the security environment used in current IIoT applications.\\
The convergence of critical infrastructure, interconnected devices, and often limited security measures within IIoT environments makes them an attractive and high-impact target for sophisticated and persistent adversaries like APT groups. These attackers seek to exploit vulnerabilities within IIoT systems to achieve their objectives, which can have significant consequences for industrial operations and, in some cases, national security. As a result, safeguarding IIoT networks from APT attacks is crucial in the realm of cybersecurity. \\
Traditional threat detection systems, including signature-based and anomaly-based approaches, face limitations in effectively detecting long-running APT campaigns \cite{chen2022machine}. Signature-based systems struggle to detect APTs that leverage zero-day exploits and new vulnerabilities \cite{garcia2009anomaly}. Conversely, anomaly-based systems, that leverage network logs \cite{yang2022systematic}, system calls \cite{breitenbacher2019hades}, and related system events \cite{xu2016sharper}, often encounter difficulties in modeling extended system behavior patterns. These systems are also vulnerable to evasion techniques since they primarily examine short sequences of system calls and events, thus limiting their ability to uncover sophisticated APT activities. \\
According to recent studies \cite{berrada2019aggregating,hassan2020tactical,barre2019mining,hossain2017sleuth}, data provenance may be a more reliable data source for identifying APTs. Data provenance depicts the flow of information between system entities, such as processes, and objects, such as files and sockets, as a directed acyclic graph (DAG), shows how a system is being used. Even when events are separated in time, this representation links the graph's causally connected events. Consequently, despite APT-affected systems often mimicking normal system behavior, the wealth of contextual information inherent in provenance data enhances the ability to distinguish between benign and malicious events \cite{zipperle2022provenance}.
Despite the demonstrated efficacy of utilizing provenance data in detecting APT attacks, researchers in the field encounter a significant challenge: the scarcity of available datasets. Moreover, the existing datasets frequently do not cover APT scenarios in IIoT environments, making it even more challenging to explore this problem. In addition to that, the APT detection methods currently proposed often lack compatibility with some features of the ever-changing APT landscape.\\
In this research, we introduce CICAPT-IIoT\footnote{Canadian Institute for Cybersecurity Advanced Persistent Threats Dataset for IIoT}, an APT attack dataset developed within an IIoT environment, to assist researchers in security analysis and developing detection methods. To achieve this, an IIoT testbed was established in a semi-controlled setting, mirroring real-world industrial operations. A realistic APT scenario, containing key APT phases like data exfiltration and defense evasion, was then implemented and executed. Raw and processed data collected during this scenario are also made available, enabling researchers to utilize and derive new features for enhanced security insights on using provenance data for the APT detection task.\\
The main contributions of our research are as follows: 
\begin{itemize}
    \item We introduce the CICAPT-IIoT dataset, a novel and comprehensive APT attack dataset captured within the IIoT environment. This dataset is generated using a hybrid testbed consisting of real and simulated IIoT components to demonstrate the complexity and diversity of modern technology systems. It contains both network logs and the provenacne data;
    \item The dataset contains more than 20 distinct attack techniques divided into eight main attack tactics that map into the APT attack scenarios, inspired by the APT29 \cite{mitreG0016} campaigns. This APT scenario enhances the dataset's effectiveness in APT detection research.
\end{itemize}

The rest of this paper is organized as follows. We discussed the related works in Section \ref{sec:relatedWorks}. Section \ref{sec:SimulatedAPTDataset} provides an overview of the testbed, its different components, and the APT attack emulation plan. we also explain the dataset generation experiments and the dataset properties in this section. Furthermore, a thorough analysis of the dataset is presented in the section \ref{sec:dataset_assess}. Finally, Section \ref{sec:conclusion} presents the conclusion of this research.

\section{Related Works}
\label{sec:relatedWorks}
Recent research has explored the utility of provenance data across various domains, including security, reproducibility, data trustworthiness, and intrusion detection \cite{lahmar2017h}. These studies have demonstrated its potential for improving the dependability and security of information systems against various threats such as APTs \cite{michael2022provenance}. Datasets play a crucial role in any attack detection research, enabling the study and modeling of behavior to identify attack activities \cite{stojanovic2020apt}. However, the majority of available datasets are generated for conventional intrusion detection rather than detecting Advanced Persistent Threats. Such datasets often lack the complexity inherent in APT cycle phases and typically comprise only network or system logs. In this section, we provide an overview of the datasets currently used in the literature for APT detection and general attack detection. Additionally, we offer a brief review of the literature on provenance data, methods for capturing provenance, and provenance-based attack detection techniques.
\subsection{Related Datasets}

The significance of datasets in attack detection research cannot be overstated, as they are fundamental to the development, testing, and refinement of detection algorithms. High-quality datasets provide a realistic representation of valuable data such as system logs and network traffic, and include both benign activities and malicious attacks, that are crucial for training and evaluating intrusion detection systems.\\
The TON IoT dataset\cite{alsaedi2020ton_iot} includes telemetry data from IoT/IIoT services and contains network traffic collected from a realistic representation of a medium-scale network at an IoT Lab. This dataset includes a variety of cyberattacks, including scanning attacks, Denial of Service (DoS) attacks, ransomware, and Man-In-The-Middle (MITM) attacks, among others, providing researchers with a comprehensive resource to study, understand, and develop countermeasures against these threats. The dataset is designed for multi-classification problems, incorporating labels for normal and attack classes, and sub-classes of attacks targeting IoT/IIoT applications.
DAPT 2020 \cite{myneni2020dapt}, is a benchmark dataset specifically designed to address the challenges in modeling and detecting APTs. This dataset includes attacks that are hard to distinguish from normal traffic flows and encompass both public-to-private interface traffic and internal network traffic. The APT stages that this dataset covers are Reconnaissance,  Foothold Establishment, Lateral Movement, and Data Exfiltration which are all crucial steps in APT campaigns. 
X-IIoTID \cite{al2021x} is a dataset for intrusion detection in the Industrial Internet of Things (IIoT) environment. IIoT systems, due to their vast connectivity and deployment of various protocols and devices, present significant security challenges. This dataset is designed to be both connectivity-agnostic and device-agnostic, thereby suitable for the heterogeneous and interoperable nature of IIoT environments. The authors state that X-IIoTD covers the Reconnaissance, Weaponization, C\&C, and Lateral movement stages of an attack scenario. 
Edge-IIoTset \cite{ferrag2022edge}, is a cybersecurity dataset designed for IoT and IIoT applications, useful for both centralized and federated learning intrusion detection systems. The dataset is generated from a custom-built IoT/IIoT testbed incorporating a wide range of devices, sensors, protocols, and cloud/edge configurations. Edge-IIoTset includes over 10 types of IoT devices that generate various types of data, such as temperature, humidity, and ultrasonic sensor readings, and contains data related to DoS, DDoS, MitM, Reconnaissance, and malware attacks. 
The DARPA OpTC dataset \cite{OpTCData} contains data from a pilot study aimed at testing the scalability of DARPA Transparent Computing technologies for cyber defense. This dataset, generated during a two-week evaluation in a highly instrumented environment, capturing both benign activities and malware injections across one thousand Windows 10 endpoints and serves as a critical resource for analyzing the effectiveness of scaled cyber defense technologies in detecting APTs within large-scale network environments.
CICIoT2023 \cite{neto2023ciciot2023} is an IoT attack dataset designed to aid in the development of security analytics applications for real IoT operations by executing 33 attacks within an IoT topology of 105 devices, classifying these attacks into seven categories: DDoS, DoS, Recon, Web-based, brute force, spoofing, and Mirai, all executed by malicious IoT devices targeting other IoT devices.
Unraveled \cite{myneni2023unraveled}, one of the most recent datasets is a semi-synthetic dataset crafted to emulate APT attacks. In response to the scarcity of publicly accessible APT datasets, the creators endeavored to enrich this dataset with a range of sophisticated attack scenarios derived from the MITRE ATT\&CK database. Additionally, they designed an Employee Behavior Generation model aimed at replicating typical employee activities. The dataset is collected during a 6-week period and contains data from Reconnaissance, Foothold Establishment, Lateral movement, and Data exfiltration stages of APTs.\\

\subsection{Provenance Data and Provenance-based Attack Detection}
Data provenance refers to the documentation or record of the origin, lineage, and history of data. It includes every step of the creation, modification, and evolution of data over time \cite{herschel2017survey}. Tracking the origins of data, such as where and how it was created and its evolution through various processing stages and transfers, are components of data provenance. 
Bates et al. \cite{bates2015trustworthy} introduced the Linux Provenance Modules (LPM), a framework designed for secure provenance collection on Linux operating systems. 
CamFlow \cite{pasquier2017practical}, using a similar architecture, implemented a practical whole-system provenance system. This system leverages the Linux Security Module (LSM) and NetFilter hooks, capturing provenance data within the Linux environment.
Some researchers developed cross-platform data provenance platforms that can collect provenance on different operating systems. For example SPADE \cite{gehani2012spade} is a provenance solution capable of tracking and analyzing provenance from multiple possibly distributed sources including OS's auditing mechanisms.There has also been some research to introduce the concept of provenance to the IoT world and use its benefits to mitigate IoT security challenges \cite{hu2020survey}. Researchers in \cite{aman2017secure,aman2019data,nwafor2017towards,sadineni2022provnet} have proposed various methods to collect and use provenance data in the IoT environment.\\
Since provenance data provides a thorough history and origin of any information within a system, it has become a valuable tool in intrusion detection systems. 
Cybersecurity researchers have explored provenance potential to improve security systems \cite{pan2023data}. Works such as UNICORN \cite{han2020unicorn} and ANUBIS \cite{anjum2022anubis} have utilized provenance data to train anomaly detection models for identifying APT activities within target environments.

\section{CICAPT-IIoT - A Semi-Synthetic IIoT APT Dataset} 
\label{sec:SimulatedAPTDataset}

In this section, we present an overview of our data collection setup and the main components of our system. As APT detection research often suffers from the lack of realistic, open-source datasets, our research involves developing CICAPT-IIoT, a semi-synthetic dataset that imitates the characteristics of APT behaviors. The dataset generation design comprises diverse tools and devices, making it possible to gather a suitable dataset for APT detection within IIoT systems.
Here we describe the data collection procedure and the various phases of the data generation process. Next, we explore our attack emulation plan and its different steps. Finally, we analyze the dataset and discuss the techniques we've used to distinguish between malicious and benign data.
\subsection{System Overview}

\begin{figure*}[htbp]
    \centering
    \includegraphics[width=\textwidth]{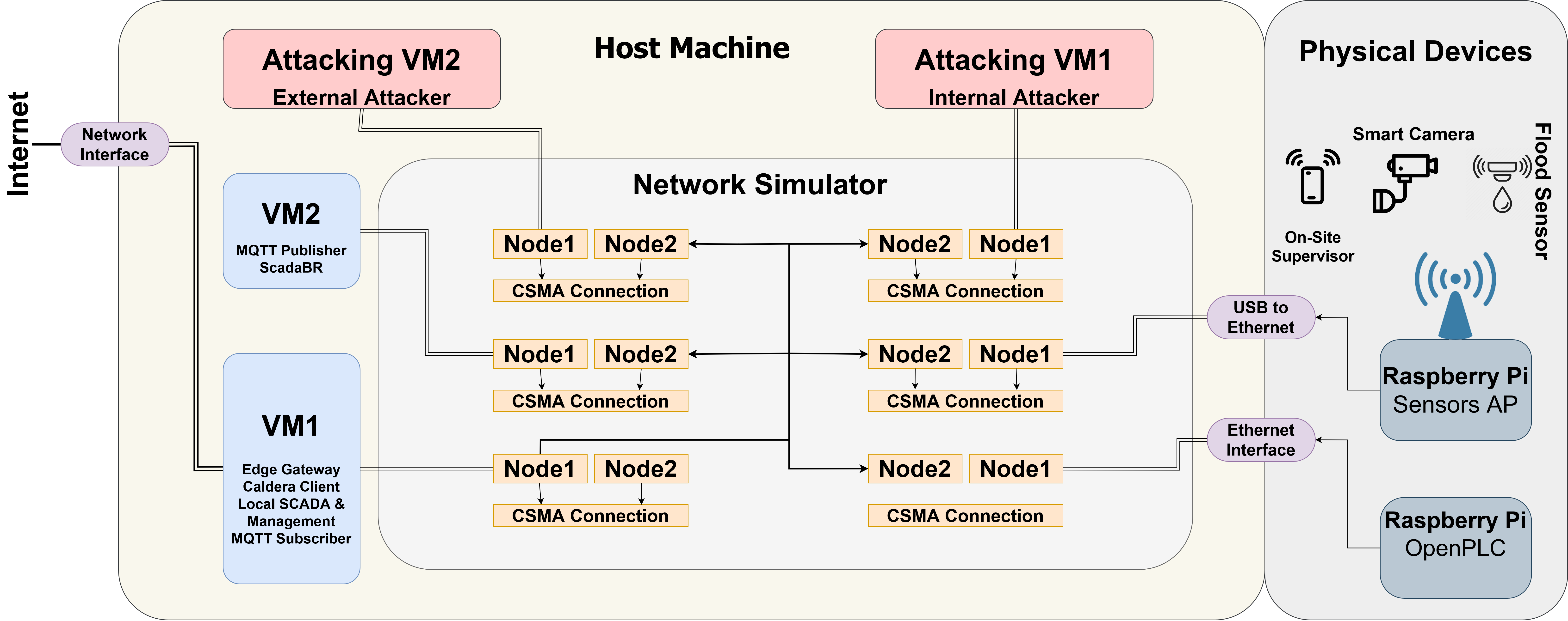}
    \caption{Overview of the testbed}
    \label{fig:Low-Level}
\end{figure*}

We have developed a simulation testbed to have a controlled environment that is useful for IIoT research and particularly beneficial for simulating APT scenarios. This testbed is based on the architecture of the Brown-IIoTbed framework \cite{al2020developing}. Our testbed's structure integrates various virtual and physical components to mirror the complexity and interactions of real-world IIoT systems. Figure \ref{fig:Low-Level} illustrates an overview of the system.\\
At the heart of our testbed is the NS3 network simulator \cite{NS3}, running on an Ubuntu host. The NS3 is essential in bridging actual and simulated nodes through its tap bridge module, allowing us to coordinate a seamless integration between real and virtual network components. 
The testbed setup involves two Ubuntu VMs and two Kali Linux VMs hosted on a machine running NS3. Ubuntu VM1 acts as the gateway, managing network traffic, and is equipped with tools like Auditd and SPADE for logging and translating logs into provenance data, respectively. It also subscribes to MQTT topics, playing a critical role within the testbed's MQTT ecosystem. Ubuntu VM2 functions as an MQTT publisher and SCADA system via SCADABR software, interacting with a PLC simulated on Raspberry Pi1, which operates with OpenPLC and communicates using the Modbus protocol. Raspberry Pi2 serves as a WiFi access point, enhancing the network's connectivity to IoT sensors. Kali VM1 and VM2 are set up as internal and external threat actors, equipped with MITRE Caldera and various attack tools, respectively. This sophisticated simulation environment not only generates comprehensive data including system logs, network traffic, and provenance data essential for APT research in IIoT environments but also mimics real IIoT operations to facilitate advanced IIoT security research. Adding devices like cameras, leakage sensors, and PLCs and using protocols such as MODBUS and MQTT helps the testbed more accurately replicate an IIoT environment \cite{jaloudi2019communication}. PLCs enable realistic automation and control simulations typical in industrial settings, and MODBUS and MQTT are common device-to-device communication protocols in IIoT systems. Table \ref{tab:testbedDevices} shows the testbed components and their roles in the experiments.

\begin{table*}[htbp]
\centering
\caption{List of the testbed components }
\begin{tabular}{|l|l|}
\hline
Device & Role \\ \hline
Ubuntu VM1 & \begin{tabular}[c]{@{}l@{}}Gateway\\ Local Management\\ MQTT Subscriber\end{tabular} \\ \hline
Ubuntu VM2 & \begin{tabular}[c]{@{}l@{}}MQTT Publisher\\ ScadaBR\end{tabular} \\ \hline
Kali VM1 & \begin{tabular}[c]{@{}l@{}}Internal Attacker\\ Caldera Server\end{tabular} \\ \hline
Kali VM2 &External Attacker \\ \hline
Raspberry Pi1 &  OpenPLC \\ \hline
Raspberry Pi2 &  WiFi Access Point \\ \hline
Litokam Smart Camera   & Camera \\ \hline
ConnectifyFlood Sensor & Flood Sensor \\ \hline
\end{tabular}
\label{tab:testbedDevices}
\end{table*}

\subsection{Attack Emulation Plan}
\label{sec:attackPlan}

\begin{table*}

    \caption{APT attack phases and techniques used in the dataset}
    \centering
    \begin{tabularx}{\textwidth}{|l|c|X|c|}
    \hline
    Tactic & Technique ID & Attack Type & APT Groups \\
    \hline
\multirow{5}{*}{Collection} & T1074 & Data Staged: Local Data Staging & APT28, APT29, APT39, APT3 \\
    \cline{2-4}
    & T1005 & Data from Local System & Andariel, APT28, APT29 \\
    \cline{2-4}
    & T1119 & Automated Collection & APT1, APT28, Chimera \\
    \cline{2-4}
    & T1113 & Screen Capture & APT28, APT39, Carbanak\\
    \cline{2-4}
    & T1115 & Clipboard Data & APT29, APT29, APT38\\
    \hline
    \multirow{2}{*}{Exfiltration} & T1560 & Archive Collected Data: Archive via Utility& APT28, APT29, APT32 \\
    \cline{2-4}
    & T1041 & Exfiltration Over C2 Channel & Lazarus, APT3, APT32 \\
    \hline
    Command and Control & T1105 & Ingress Tool Transfer & Lazarus, APT29, APT3 \\
    \hline
    \multirow{2}{*}{Persistence} & T1546 & Event Triggered Execution  & APT28, APT29, APT3 \\
    \cline{2-4}
    & T1136 & Create Account: Local Account  & Dragonfly, FIN13, APT29 \\
    \hline
    \multirow{9}{*}{Discovery} & T1087 & Account Discovery: Local Account & APT1, APT3, Chimera  \\
    \cline{2-4}
    &\multirow{2}{*}{T1016} & System Network Configuration Discovery: Internet Connection Discovery & FIN13, Gamaredon, APT29  \\
    
    \cline{3-4}
    && System Network Configuration Discovery: Wi-Fi Discovery & Magic Hound, Wizard Spider \\
    \cline{2-4}
    & T1033 & System Owner/User Discovery  & Chimera, Dragonfly, APT3 \\
    \cline{2-4}
    & T1518 & Software Discovery & HEXANE, MuddyWater  \\
    \cline{2-4}
    & T1069 & Permission Groups Discovery: Local Groups & Chimera, HEXANE, APT29  \\
    \cline{2-4}
    & T1082 & System Information Discovery  & Chimera, APT3, APT32 \\
    \cline{2-4}
    & T1083 & File and Directory Discovery  & APT28, APT29, APT32 \\
    \cline{2-4}
    & T1018 & Remote System Discovery  & Chimera, APT29, APT32 \\
    \hline
    \multirow{3}{*}{Credential Access} & \multirow{2}{*}{T1552} & Unsecured Credentials: Credentials In Files  & APT3, APT33, FIN13 \\
    \cline{3-4}
    && Unsecured Credentials: Bash History  & - \\
    \cline{2-4}
    & T1555 & Credentials from Password Stores: Credentials from Web Browsers  & APT33, APT39, HEXANE\\
    \hline
    Lateral Movement & T1021 & Remote Services: SSH  & APT29, APT39, Lazarus\\
    \hline
    \multirow{2}{*}{Defense Evasion} & T1036 & Masquerading: Right-to-Left Override  & APT28, APT29, Dragonfly \\
    \cline{2-4}
    & T1485 & Data Destruction  & APT38, Gamaredon, Lazarus\\
    \hline
    \end{tabularx}
    \label{tab:attack_phases}
\end{table*}

APT29, also known as `The Dukes' or `Cozy Bear', is a sophisticated cyber threat group noted for its advanced cyber espionage tactics and persistent attacks. The group's activities have had significant impacts, leading MITRE to publish an adversary emulation plan for APT29 within the framework of MITRE Caldera \cite{mitreAPT29}. This emulation plan, however, uses tactics and techniques mostly tailored for Windows environments, which are not directly transferable to Linux systems. Therefore, we  adapted APT29 emulation plan and the MITRE ATT\&CK framework's Tactics, Techniques, and Procedures (TTPs)  
to design a customized attack emulation plan suitable for the Linux-based testbed. This plan aims to replicate APT29’s operational patterns within the unique environment of the testbed. Table~\ref{tab:attack_phases} shows the different APT tactics and techniques used in the dataset. The developed emulation plan encompasses several stages, each reflecting a typical phase in an APT campaign, including Data Collection and Exfiltration, Deployment of Stealth Toolkits for further activities, Defense and Discovery Evasion, Maintaining Persistence, Accessing System Credentials, and Lateral Movement to other components in the network.

\subsection{Data Collection and Experiments}
The NS3 simulator serves as the primary platform for our testbed, managing network connections and enabling the monitoring and logging of all network packets. It operates in Real-Time mode to facilitate the integration of real and simulated nodes. Along with network logs collected by NS3, system logs are gathered using the Linux Audit Daemon (Auditd), which leverages the Linux Auditing System for efficient log capture. These logs are processed by SPADE \cite{gehani2012spade}, a service that generates provenance data, enriching the dataset with an additional layer of information useful in APT detection.\\
The experiments were conducted in two phases. The first phase, conducted over four days, simulated normal system operations to establish a baseline behavior of the testbed components including VMs, sensors, and Raspberry Pis. The second phase, spanning three days, simulated an APT attack using APT29 tactics executed through Kali VM1 with MITRE Caldera. This phase followed the APT's 'low and slow' approach as attack steps are executed in random time intervals of 45 to 75 minutes to mimic the stealth and persistence typical of APTs and closely replicate real-world attack dynamics.

\subsection{Dataset Properties}
The dataset is organized into two folders: phase1 data and phase2 data, each containing two types of data—provenance data and network packets. The provenance data files are in CSV format and contain the nodes and edges of the provenance graph. Each node in the provenance data is assigned a unique 32-digit ID, which is utilized by the edge entries to establish connections between nodes in the graph.\\

\begingroup\small
\begin{longtable}[c]{|l|l|l|l|}
\hline
\textbf{\# } & \textbf{Feature}        & \textbf{Description}               & \textbf{Provenance Type}            \\ \hline
\endhead

\cellcolor{gray!25}1  & \cellcolor{gray!25}id             & \cellcolor{gray!25}Node identifier       & \cellcolor{gray!25}All node types             \\ \hline
2  & type           & Edge or node type      & All nodes and edges        \\ \hline
\cellcolor{gray!25}3  & \cellcolor{gray!25}from           & \cellcolor{gray!25}Source node ID            & \cellcolor{gray!25}Edges                      \\ \hline
4  & to             & Destination node ID       & Edges                      \\ \hline
\cellcolor{gray!25}5  & \cellcolor{gray!25}uid            & \cellcolor{gray!25}User Id                   & \cellcolor{gray!25}Process nodes              \\ \hline
6  & egid           & Effective group ID        & Process nodes              \\ \hline
\cellcolor{gray!25}7  & \cellcolor{gray!25}exe            & \cellcolor{gray!25}Executable path           & \cellcolor{gray!25}Process nodes              \\ \hline
8  & gid            & Group ID                  & Process nodes              \\ \hline
\cellcolor{gray!25}9  & \cellcolor{gray!25}euid           & \cellcolor{gray!25}Effective user ID         & \cellcolor{gray!25}Process nodes              \\ \hline
10 & name           & Executable name           & Process nodes              \\ \hline
\cellcolor{gray!25}11 & \cellcolor{gray!25}pid            & \cellcolor{gray!25}Process ID                & \cellcolor{gray!25}Process nodes, WDF edges   \\ \hline
12 & seen time      & Process seen time         & Process nodes              \\ \hline
\cellcolor{gray!25}13 & \cellcolor{gray!25}source         & \cellcolor{gray!25}Data origin               & \cellcolor{gray!25}All nodes and edges        \\ \hline
14 & ppid           & Parent process ID         & Process nodes              \\ \hline
\cellcolor{gray!25}15 & \cellcolor{gray!25}command line   & \cellcolor{gray!25}Full command line used    & \cellcolor{gray!25}Process nodes              \\ \hline
16 & start time     & Process start time        & Process nodes              \\ \hline
\cellcolor{gray!25}17 & \cellcolor{gray!25}event ID       & \cellcolor{gray!25}Unique event ID           & \cellcolor{gray!25}All edges                  \\ \hline
18 & time           & Event time                & All edges                  \\ \hline
\cellcolor{gray!25}19 & \cellcolor{gray!25}operation      & \cellcolor{gray!25}Type of operation         & \cellcolor{gray!25}All edges                  \\ \hline
20 & path           & File path                 & File,link, directory nodes \\ \hline
\cellcolor{gray!25}21 & \cellcolor{gray!25}subtype        & \cellcolor{gray!25}Subtype of nodes          & \cellcolor{gray!25}Artifact nodes             \\ \hline
22 & permissions    & Access permissions        & File,link, directory nodes \\ \hline
\cellcolor{gray!25}23 & \cellcolor{gray!25}epoch          & \cellcolor{gray!25}Sequence number           & \cellcolor{gray!25}Artifact nodes             \\ \hline
24 & version        & Version number            & Artifact nodes             \\ \hline
\cellcolor{gray!25}25 & \cellcolor{gray!25}Flag           & \cellcolor{gray!25}Resource access mode      & \cellcolor{gray!25}Used, WGB edges            \\ \hline
26 & remote port    & Port number               & Network socket nodes       \\ \hline
\cellcolor{gray!25}27 & \cellcolor{gray!25}protocol       & \cellcolor{gray!25}Used protocol             & \cellcolor{gray!25}Network socket nodes       \\ \hline
28 & remote address & IP address                & Network socket nodes       \\ \hline
\cellcolor{gray!25}29 & \cellcolor{gray!25}tgid           & \cellcolor{gray!25}Thread group ID           & \cellcolor{gray!25}Unknown nodes              \\ \hline
30 & fd             & File descriptor           & Unknown nodes              \\ \hline
\cellcolor{gray!25}31 & \cellcolor{gray!25}mode           & \cellcolor{gray!25}Permission setting        & \cellcolor{gray!25}WGB edges                  \\ \hline
32 & label          & Node label- attack/benign & All nodes                  \\ \hline
\cellcolor{gray!25}33 & \cellcolor{gray!25}subLabel       & \cellcolor{gray!25}Attack category           & \cellcolor{gray!25}All nodes                  \\ \hline
\caption{Provenance Data Features}
\label{tab:Provenance_Features}
\end{longtable}
\endgroup

Besides the IDs, the provenance data files comprise 32 features in total. However, due to the heterogeneous nature of nodes and edges that are all in a single file, not all features apply to every node or edge type, resulting in many fields being populated with NaN values. Table \ref{tab:Provenance_Features} lists all features provided in the provenance data part of the dataset. The provenance data includes two main node types: Process and Artifact. The Artifact node type is further categorized into various subtypes such as file, directory, network socket, link, and unknown, the latter being used for provenance node types that do not fit into the existing subtypes. 
The other data type in the dataset is the network logs captured using NS3 during the experiments and stored in pcap format. These pcap files can be further processed into CSV format. We generate the CSV format from these pcaps that have the information at the packet level and contain 67 features for each packet. The last file in the dataset is the Attack Information file, which contains all necessary information about the attacks performed during the experiments in phase 2. This information includes attack time, attack PID, and the category of attack. This file helps the researchers to further analyze the dataset behavior during the attacks.

\begin{figure*}[htbp]
    \centering
    \includegraphics[width=\textwidth]{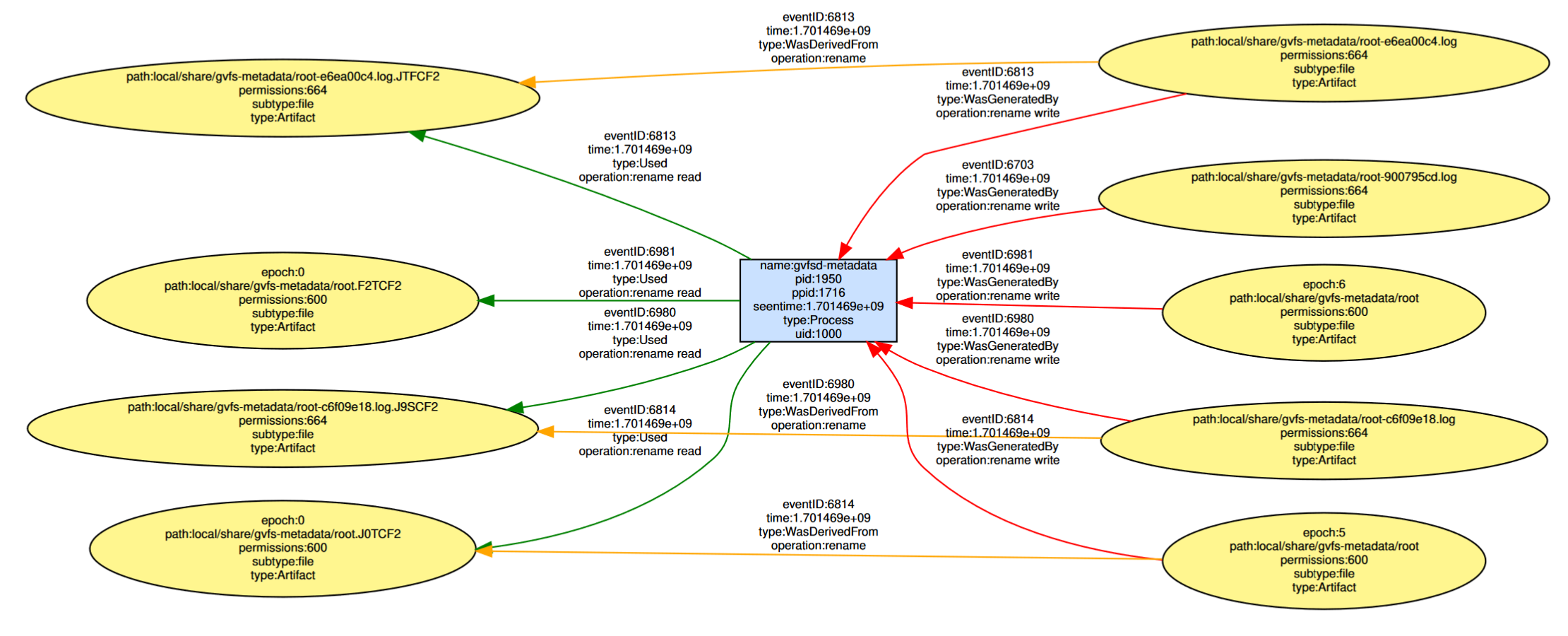}
    \caption{An example of provenance graph}
    \label{fig:sample}
\end{figure*}

Figure \ref{fig:sample} displays an example of a provenance graph. 
This is a subgraph of the complete provenance graph based on the the CICAPT-IIoT dataset. This provenance graph shows several rename operations performed on files, resulting in the generation of a new set of files.

\section{Dataset Assessment}
\label{sec:dataset_assess}
\subsection{General Statistics of the Dataset}
The dataset was generated in two phases: Phase 1, which lasted approximately 96 hours, and Phase 2, which lasted about 72 hours. It contains of about 10 GB of data. During both phases, network logs and system logs were collected, and provenance logs were generated using the system logs and SPADE. A detailed breakdown of the dataset's segments can be found in Table \ref{table:data_distribution}. 
As shown in the Table \ref{table:data_distribution}, the CICAPT-IIoT dataset is notably unbalanced, with approximately 99.5\% of the samples representing normal behavior and only a small fraction indicating malicious activities, which is typical in APT scenarios. This significant imbalance is reflective of real-world conditions in IIoT environments, where actual attacks are infrequent relative to regular operations.\\
In such contexts, using oversampling techniques to artificially balance the dataset—by replicating the minority class or generating synthetic samples—can be counterproductive. Although these methods might facilitate algorithmic training in the short term, they distort the reality of how APTs manifest within network systems. By oversampling attack data, the models are trained on a scenario that is not reflective of actual operational conditions. This training approach can lead to models that perform well on balanced or altered datasets in testing environments but fail to detect genuine APT activities when deployed in real-world scenarios.

\begin{table}[htbp]
\small
\centering
\caption{Data distribution across different phases and data types}
\begin{tabular}{|l|l|c|c|}
\hline
\textbf{Attribute} & \textbf{Event type} & \textbf{Provenance Data} & \textbf{Network Data} \\ \hline
\textbf{Phase 1} & \textbf{Benign} & 46773 Nodes & 12103705 Packets \\ \hline
\multirow{10}{*}{\textbf{Phase 2}} 
& \textbf{Benign} & 52954 Nodes & 9535819 Packets \\ \cline{2-4}
& \textbf{Attack} & 330 & 1004 \\ \cline{2-4}
& \textbf{Collection} & 100 & 460 \\ \cline{2-4}
& \textbf{Exfiltration} & 22 & 42 \\ \cline{2-4}
& \textbf{Credential Access} & 82 & 58 \\ \cline{2-4}
& \textbf{Defence Evasion} & 45 & 192 \\ \cline{2-4}
& \textbf{Discovery} & 36 & 138 \\ \cline{2-4}
& \textbf{Persistence} & 19 & 44 \\ \cline{2-4}
& \textbf{C\&C} & 16 & 56 \\ \cline{2-4}
& \textbf{Lateral Movement} & 10 & 14 \\ \hline
\end{tabular}
\label{table:data_distribution}
\end{table}



\subsection{Comparison Against Similar Datasets}

\begin{table*}[htbp]
\caption{Related datasets analysis}
\centering
\begin{tabularx}{\textwidth}{@{}lccccccc@{}}
\toprule
Dataset & \makecell{Edge-IIoTset \\ \cite{ferrag2022edge}} & \makecell{CICIoT2023\\\cite{neto2023ciciot2023}} & \makecell{X-IIoTID\\\cite{al2021x}} & \makecell{TON IoT\\\cite{alsaedi2020ton_iot}}& \makecell{OpTC\\\cite{OpTCData}} & \makecell{DAPT 2020\\\cite{myneni2020dapt}} & \makecell{CICAPT IIoT \\(This Work)} \\ \midrule
IoT/IIoT & \checkmark & \checkmark & \checkmark & \checkmark & & &\checkmark \\
Network Logs & \checkmark & \checkmark & \checkmark & \checkmark & & \checkmark &\checkmark \\
Provenance/Host Logs & & & & & \checkmark & &\checkmark \\
Duration &N/A&16 Hours&N/A&N/A&7 Days&5 Days&7 days\\
\hline
Establish foothold & \checkmark & \checkmark & & \checkmark & & \checkmark &\checkmark \\
Collection & & & & & & &\checkmark \\
Data exfiltration  & & & \checkmark & & \checkmark & \checkmark &\checkmark \\
Command \& Control & \checkmark & \checkmark & \checkmark &  & \checkmark & &\checkmark \\
Persistence & \checkmark & \checkmark & & \checkmark & \checkmark & &\checkmark \\
Discovery & \checkmark & \checkmark & \checkmark & \checkmark & \checkmark & \checkmark &\checkmark\\
Credential Access & \checkmark & \checkmark & \checkmark & \checkmark & \checkmark& &\checkmark \\
Lateral movement & & & \checkmark & & \checkmark & \checkmark &\checkmark \\
Defence Evasion & & & & & \checkmark& &\checkmark \\ \bottomrule
\end{tabularx}
\label{tab:datasets}
\end{table*}

The CICAPT-IIoT dataset stands out from other datasets in several ways. 
First, the inclusion of multiple data sources enhances the analytical capabilities of researchers, and supports the development of new detection methods that utilize both network data and provenance logs. Furthermore, as APT attacks are known for their multi-stage operations, they require a comprehensive coverage of all associated stages, tactics, and techniques to effectively model APT campaigns in a cybersecurity dataset. Many existing datasets either do not directly address all APT tactics, as they only map network-based attacks to APT stages, or they fail to cover all the stages necessary to fully represent an APT campaign. In contrast, the CICAPT-IIoT dataset aims to provide a complete and realistic portrayal of an APT attack, encompassing the most relevant and authentic stages and techniques.
Table \ref{tab:datasets} provides an analysis of some of the related datasets. The comparison of these datasets is based on several key factors: the environment in which the data was collected, the types of data included, and the APT tactics they cover. 

\section{Conclusion}
\label{sec:conclusion}
Given the escalating threat of APT attacks on IIoT systems, developing effective detection solutions is crucial. Datasets are central to these efforts as they enable the development of defenses against such sophisticated threats. In this paper, we present CICAPT-IIoT, a dataset designed for IIoT environments, aimed at helping researchers in security analysis and the design of detection techniques against APTs. The dataset contains over 20 well-known attack techniques, forming 8 different tactics commonly utilized in APT campaigns. The collected data in provenance and network log formats are available\footnote{ CIC website: \url{https://www.unb.ca/cic/datasets/index.html}}\\

\bibliographystyle{unsrt}  
\bibliography{references}  

\end{document}